\documentclass[twoside]{dis04}
\def\runauthor{E.Avetisyan et al}
\def\shorttitle{Beam-Spin Azimuthal Asymmetries at HERMES}
\begin{document}

\title{Beam-Spin Azimuthal Asymmetries \\
in Pion Electroproduction at HERMES
}

\author{\underline{Eduard Avetisyan}$^{1,2}$, Armine Rostomyan$^2$, Alexander Ivanilov$^3$ \\
(on behalf of The HERMES collaboration) }

\address{  \footnotesize\it
$^1)$ Laboratori Nazionali di Frascati, INFN, 00044 Frascati (Rome), Italy \\
E-mail: Eduard.Avetisyan@desy.de \\
 \footnotesize\it
 $^2)$ Yerevan Physics Institute, Yerevan 375036, Armenia \\
 \footnotesize\it
 $^3)$ Institute for High Energy Physics, Protvino 142281, Russia \\
}

\maketitle

\abstracts{We present the measurement of Single Beam-Spin Azimuthal Asymmetries
in pion electroproduction off hydrogen in semi-inclusive deep inelastic
scattering (SIDIS). The measurement was made
using the HERMES spectrometer with an internal gas
target and the polarized 27.6 GeV electron(positron) beam of HERA 
using the data taken during the years 1996-2000.
The  $\sin\phi$ modulation of the azimuthal asymmetry was measured for 
semi-inclusive $\pi^+,\pi^-$ and $\pi^0$. The dependence of
the asymmetry on the Bjorken $x$, pion relative ($z$) and transverse ($P_T$)
momentum is presented. Results are compared to theoretical model calculation.
}

\section{Introduction}

Single-spin asymmetries (SSA) in SIDIS are known as a powerful tool for probing the 
 spin structure of the nucleon.  They give access to 
transverse momentum dependent (TMD) \cite{TANG,BELJI,JIMA} parton distribution and
naive time-reversal-odd (T-odd) fragmentation functions with
  prominent examples like the transversity distribution $h_1$ \cite{JAFFE}, Sivers $f_{1T}^\perp$
\cite{Sivers} and Collins $H_1^\perp$ \cite{COL} functions.
Significant SSA were observed with
longitudinally polarized target \cite{AUL}, transversely polarized
target \cite{TRANS}  and longitudinally polarized beam
\cite{JLAB}.
Beam-spin asymmetries (BSA) receive contributions from different combinations of twist-2 and
twist-3  \cite{LM,KOGAN,YUAN,BAC} TMD functions and in particular may provide 
access to the Collins function, which is important for the
ongoing transversity measurements at HERMES and COMPASS.
In this contribution we report measurements of beam SSA for
$\pi^+,\pi^-$ and $\pi^0$ using the HERMES detector at HERA.

\section{The Semi-Inclusive Deep Inelastic Scattering Cross Section}
Assuming that factorization holds in quark scattering and  fragmentation processes
the SIDIS cross section can be given as a 
convolution of a distribution function $f^{H\rightarrow q}$ (DF), an elementary 
hard scattering process cross section, and a fragmentation function $D^{q\rightarrow h}$ (FF),
\begin{equation}
\sigma^{eH\rightarrow ehX} = {f^{H\rightarrow q}} \otimes 
\sigma^{eq\rightarrow eq} \otimes D^{q\rightarrow h}
\label{conv}
\end{equation} 

\begin{figure}[h]
\begin{minipage}{4.5cm}
\begin{eqnarray*}
Q^2  &  = &  4EE^\prime \sin^2 ( \frac{\theta}{2}) \\
\nu  &  = &  E-E^{\prime} \\
x    &  = &  \frac{Q^2}{2m\nu} \\
y    &  = &  \frac{\nu}{E} = \frac{p\cdot q}{p\cdot k}\\
z    &  = &  \frac{E_h}{\nu}  \\
\phi  &  = & {\vec q \times \vec k \cdot \vec P_h \over |\vec q \times \vec k \cdot \vec P_h|} 
\cos^{-1} {\vec q \times \vec k  \cdot \vec q \times \vec P_h \over |\vec q \times \vec k| \cdot |\vec q \times \vec P_h|}
\end{eqnarray*}
\end{minipage}
\begin{minipage}{7.5cm}
\center{ 
\includegraphics[width=6.5cm]{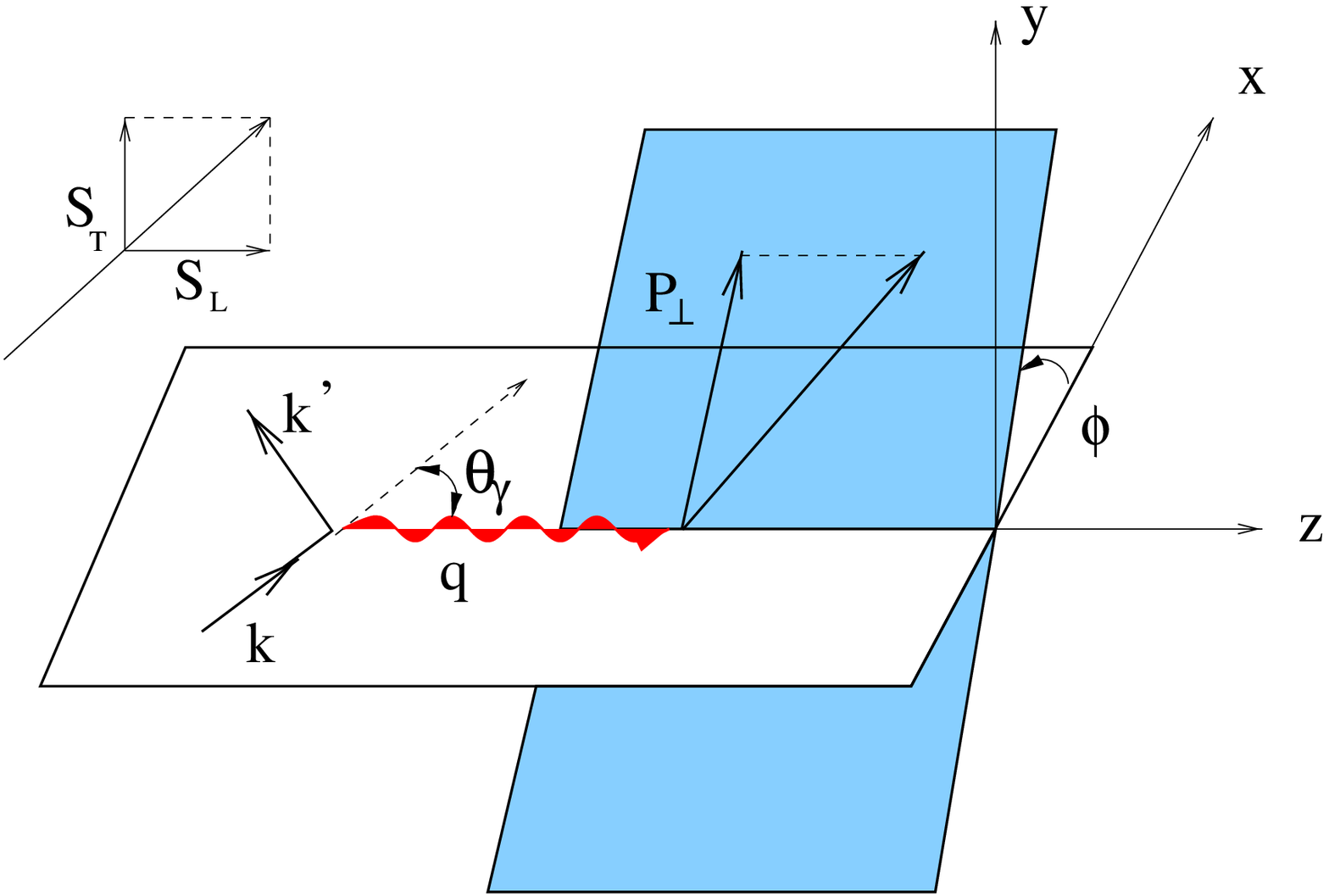}
}
\end{minipage}
\caption[*]{Definitions of kinematic planes and variables}
\label{kinem}
\end{figure}
\vspace{0.5cm}

The  cross section
in the polarized beam case can be written as \cite{BAC}:
\begin{eqnarray}
\label{siglu}
\frac{d\sigma_{LU}}{dxdydzd\phi} & = &
\lambda_e  \sin\phi\,y\sqrt{1-y} \frac{M}{Q}\sum_{a,\bar{a}} {\bf{e^2_a}} x^2  \\
& & \left[ {xze^a(x)} {H_1^{\perp a}(z)}   + {h_1^{\perp a}(x)}{E^a(z)}  
+\frac{M_h}{M} f_1^a(x) G^{\perp a}(z) - \right. \nonumber \\
& & \left. - xz\frac{M}{M_h} g^{\perp a}(x) D_1^a(z) 
-\frac{m}{M} z f_1^a(x)H_1^{\perp a}(z) - \frac{m}{M_h}z h_1^{\perp a}(x) D_1^a(z)\right] \nonumber
\end{eqnarray}
where $\lambda_e$ is the beam polarization, $m,M$ and $M_h$ are the quark,nucleon and hadron masses, 
 $Q$ is the hard scale and the azimuthal angle $\phi$ is defined as the angle
 between the lepton scattering and hadron production planes (see Fig.\ref{kinem}). 
The ``L'' 
and ``U'' indices stand for longitudinally polarized beam and unpolarized target, 
respectively. The sum is made over all quark flavors. The $f(x)$ and $D(z)$ are the
corresponding  distribution and fragmentation functions, 
respectively.
(The complete tree-level description up to twist three for various processes can be found in
\cite{LM}). 

\section{The Beam-Spin Asymmetry Measurement}

        The analyzing power for the beam-spin asymmetry can be extracted
 as corresponding moment of the event distributions for the two beam helicity
states:
\begin{equation}
A_{LU}^{sin\phi}= {1 \over |P_B|}\frac { \sum_{i=1} ^{N^\uparrow} sin\phi_i - 
 \sum_{i=1} ^{N^\downarrow} sin\phi_i } { {1\over 2} 
 (N^\uparrow + N^\downarrow)}
\label{asym}
\end{equation}
where $\uparrow /\downarrow$ denotes positive/negative helicity of the beam,
 $N^{\uparrow / \downarrow}$ is the number of selected events with a 
detected hadron for each beam spin state and $|P_B|\simeq 0.533 \pm 0.029$ is the mean absolute beam
polarization measured by the two HERA polarimeters \cite{BPOL}.
\begin{figure}[t]
\vspace*{7.0cm}
\begin{center}
\includegraphics{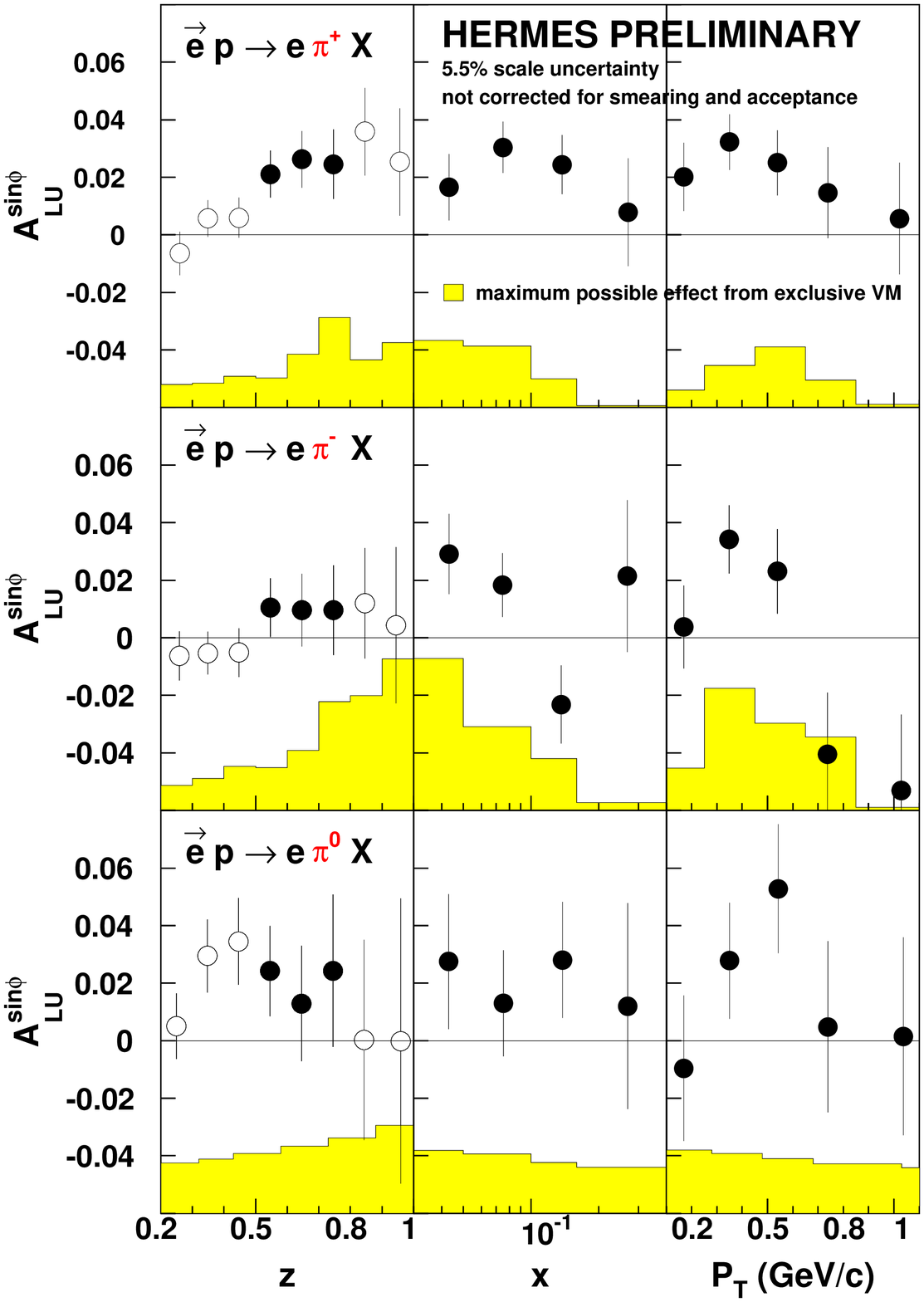}
\vspace{4cm}
\caption[*]{ Dependence of beam SSA on $z$, $x_B$ and $P_T$
for pions. The measurement of the $x_B$ and $P_T$ dependences is
made for $0.5<z<0.8$ range (indicated by full circles on the
left column).  Error bars represent the statistical uncertainty and the shaded bands
represents an upper limit for
a possible 
uncertainty from exclusive vector meson decay product contribution 
in the semi-inclusive pion sample. An additional 5.5\% scale uncertainty 
is due to the beam polarization measurement.
}
\label{alupi}
\end{center}
\end{figure}
The scattered electrons/positrons ( jointly referred to as 
positrons) and associated hadrons were 
detected by the HERMES spectrometer \cite{SPEC}.
 The separation between positrons and hadrons
was accomplished by the use of a set of particle identification detectors: a transition radiation detector,
a preshower, a threshold \v Cerenkov detector (replaced by a RICH detector \cite{RICH} since 1998) 
and 
an electromagnetic calorimeter. The average identification efficiency exceeded 98\% with
a hadron contamination in the positron sample below 1\%.
The scattered  positron was required to satisfy a set of cuts, namely $1<Q^2<15$ GeV$^2$, 
$0.023<x<0.4$, $W^2>4$ GeV$^2$, $y<0.85$. 
For identification of charged pions the \v Cerenkov and RICH detectors were used 
during the corresponding data taking periods. To assure reliable identification
of pions ($>95\%$) in both detectors the momentum range of $4.5<p<13.5$ GeV was chosen.

Neutral pions were identified requiring two trackless neutral
clusters in the electromagnetic calorimeter with energies above a threshold of 1 GeV. 
By the reconstruction of the invariant mass of two photons $M_{\gamma\gamma}$,
 a clear peak around the $\pi^0$
mass is observed with a resolution of about $0.012$ GeV. Neutral pions
have been selected requiring \mbox{$0.11<M_{\gamma\gamma}<0.16$ GeV} . 
The asymmetry of the combinatorial 
background has been measured outside the mass window 
of the $\pi^0$ peak and a background subtraction has been
 applied to account for this.

The analyzing powers $A_{LU}^{\sin\phi}$ as a function of $z$, $x$ and 
$P_T$ (transverse momentum w.r.t. virtual photon direction) are shown
in Fig.\ref{alupi}. To reduce the contamination of the events
from the target fragmentation region
and contributions from exclusive processes a range of \mbox{$0.5<z<0.8$} was chosen
in the measurements of the $x$ and $P_T$ dependences (full circles in Fig.\ref{alupi}). 
The measured
$A_{LU}^{\sin\phi}$ asymmetry
of $\pi^+$ is positive in the order of $2\%$ and has a clear rise with increasing $z$, while
the $x$ and $P_T$ dependences show a hint of a falloff in the lowest and highest bins.
This measurement extends a former
CLAS measurement of $A_{LU}^{\sin\phi}$  at JLAB \cite{JLAB}
 to lower $x$ and higher $Q^2$ regions.
 The $z$ and $P_T$ dependences from
both HERMES and CLAS measurements are
in  agreement within statistical accuracy
after accounting for differences in the kinematics. 
The $\pi^+$ 
asymmetry is in 
 agreement with a quark-diquark model based prediction  that considers 
$e(x)H_1^\perp(z)$ and $h_1^\perp(x)E(z)$ terms \cite{KOGAN}. 
Other theoretical models \cite{YUAN,SCHWEITZER} calculating  different components of
Eq.\ref{siglu} are also in qualitative agreement with our measurement.
The asymmetries measured for the first time on $\pi^-$ and $\pi^0$ 
are positive, but compatible with zero 
within the present  statistical uncertainties.

The semi-inclusive pion production  ($ep\rightarrow e^\prime \pi X$)
 with an underlying mechanism
of quark fragmentation
is diluted by exclusive 
vector meson 
production which can contribute significantly in certain kinematic ranges.
Their contribution in the semi-inclusive pion sample has been investigated
using a Pythia Monte-Carlo tuned for HERMES kinematics \cite{PATTY}.
To estimate the possible contribution to the asymmetry, the $A_{LU}^{\sin\phi}$ 
has been measured for the decay products of $\rho^0$ mesons (which  dominate
the contamination). 
Due to  lack of statistics that procedure
was not used for $\omega$ and charged $\rho$ mesons. For the latter cases a maximum possible
asymmetry was assumed, constrained by only basic principles (positivity limits, 
kinematic suppression factor and asymmetry transfer coefficients).
The resulting interpretive uncertainty is indicated by the shaded band in Fig.\ref{alupi}.

In conclusion, we presented the beam induced single spin azimuthal asymmetry 
measured in the production of single pions in semi-inclusive DIS. The measured
$A_{LU}^{\sin\phi}$
of $\pi^+$ is positive in the order of $2\%$ and has a clear rise 
with increasing $z$, while
the $x$ and $P_T$ dependences exhibit a hint of a falloff in the lowest and highest bins.

\section*{Acknowledgments} We would like to thank Harut Avakian, Delia Hasch and
Nicola Bianchi for useful discussions; Elke Aschenauer and Alexander Borissov for
assistance in Monte Carlo studies.

\end{document}